**Atomic mechanisms for the Si atom dynamics in graphene: chemical transformations at the edge and in the bulk**

Maxim Ziatdinov[1-3], Ondrej Dyck[1,2], Stephen Jesse[1,2], and Sergei V. Kalinin[1,2]

Affiliations:

[1] The Center for Nanophase Materials Science, Oak Ridge National Laboratory, Oak Ridge, TN

[2] The Institute for Functional Imaging of Materials, Oak Ridge National Laboratory, Oak Ridge, TN

[3] Computational Sciences & Engineering Division, Oak Ridge National Laboratory, Oak Ridge, TN

Recent advances in scanning transmission electron microscopy (STEM) allow to observe solid-state transformations and reactions in materials induced by thermal stimulus or electron beam on the atomic level. However, despite the rate at which large volumes of data can be generated (sometimes in the gigabyte to terabyte range per single experiment), approaches for the extraction of material-specific knowledge on the kinetics and thermodynamics of these processes are still lacking. One of the critical issues lies in being able to map the evolution of various atomic structures and determine the associated transition probabilities directly from raw experimental data characterized by high levels of noise and missing structural elements. Here, we demonstrate an approach based on the combination of multiple machine learning techniques to study the dynamic behavior of e-beam irradiated Si atoms in the bulk and at the edges of single-layer graphene in STEM experiments. First, a deep learning network is used to convert experimental STEM movies into coordinates of individual Si and carbon atoms. Then, a Gaussian mixture model is further used to establish the elementary atomic configurations of the Si atoms, defining the bonding geometries and chemical species and accounting for the discrete rotational symmetry of the host lattice. Finally, the frequencies and Markov transition probabilities between these states are determined. This analysis enables insight into the thermodynamics of defect populations and chemical reaction networks from the atomically resolved STEM data. Here, we observe a clear tendency for the formation of a 1D Si crystal along zigzag direction of graphene edges and for the Si impurity coupling to topological defects in bulk graphene.

Solid state transformations underpin multiple areas of condensed matter physics, materials science, and chemistry, ranging from structural transitions in ferroics, diffusionless phase transformations in metals and alloys, solid state reactions and phase separations in multicomponent systems, and a broad range of solid-gas and solid-liquid reactions that underpin the operation of batteries and fuel cells. While a broad range of macroscopic theories have been developed to describe these processes depending on the nature of transformations, ranging from stochastic Kolmogorov-Avrami type models for first-order transformations,[1–3] phase field models,[4] and more complex formalisms based on the solid state defect chemistry descriptions, remarkably little is known on the local mechanisms. These include the local atomic configurations, defect population and dynamics, and their transformation networks. The reason for this dearth of information is straightforward – while complex defect chemistry models can be developed and incorporated in the reaction-diffusion or phase-field formalisms, the number of measurable variables in macroscopic experiments is extremely limited, significantly complicating the fundamental studies of the associated mechanisms. In fact, even the nature of the structural descriptors and collective variables describing the system are generally unknown.

Significant progress in this area has been achieved with the introduction and widespread adoption of transmission electron microscopy (TEM) and, particularly, aberration corrected scanning transmission electron microscopy (STEM).[5] Early TEM studies provided significant insight into the structure and population of extended defects such as dislocations and crystallographic shear structures,[6] and enabled extensive visualization of solid-state reaction fronts. Aberration corrected STEM has enabled the imaging of materials with high propensity for beam damage and was further extended towards observations of in-situ processes induced by temperature, gas, or bias stimuli. While in 3D solids atomic level studies are often limited by the fact that reaction fronts are rarely uniform in the beam direction, the recent interest in 2D materials provides model systems in which visualization of virtually all atomic units during a reaction process is possible.[7]

Particularly of interest are dynamic phenomena induced by the effect of the electron beam during the imaging. Traditionally, these effects were perceived as undesirable beam damage, where e-beam irradiation of the sample led to phase decomposition, sputtering, and other damage.[8,9] However, advances in low-voltage probes have enabled studies in which only a small number of atomic species and chemical bonds change during imaging, enabling dynamic studies of beam-induced transformations.[10–13] Phenomena such as oxygen vacancy ordering,[14] dopant atom dynamics,[12] formation of topological defects and bond formation and breaking,[15–17] and vacancy formation[18] etc. have been visualized with atomic resolution. Particularly of interest are the combination of the electron beam induced reactions, real-time image-based feedback, and externally controlled beam motion that enabled matter sculpting and direct atom by atom manipulation,[7,19–25] a feat previously accessible only to scanning probe microscopy methods.[26,27]

However, despite the broad range of beam induced phenomena resolved with atomic resolution, approaches for the analysis of associated mechanisms are still lacking. Indeed, while the evolution of atomically resolved images with time provides great insight into the qualitative character of atomic motion, the associated mechanisms that describe the kinetics and thermodynamics of these processes via low dimensional representations are of keen interest and can be further generalized and compared to theoretical models.

Here, we develop a comprehensive approach to describe the thermodynamics and kinetics of beam induced processes from the STEM data. First, deep learning methods are used to denoise the images and extract atomic coordinates from the individual frames. Then, a gaussian linear mixture model is used to create a library of the structural descriptors. Finally, a Markov process-based framework is used to analyze the transition probabilities, providing a comprehensive analysis of the reversible electron beam-induced processes. We explore the beam induced reactions of Si atoms on the edge of graphene and in the graphene bulk. This framework is more general and can be applied to the analysis of reactions and chemical transformations in solids under specific conditions, as discussed below.

As a model system, we have chosen to examine the atomically-resolved evolution of Si dopants at the edge of a hole in graphene, driven by 60 keV electron beam irradiation (Figure 1(a)). A second experiment captures the beam induced decomposition of an, initially, pristine area of graphene under a 100 keV beam (Figure 4(a)). In these experiments, the image acquisition itself, accomplished by scanning the beam over the sample (and recording the high angle annular detector (HAADF) signal), is the principle energy source for the observed sample evolution.

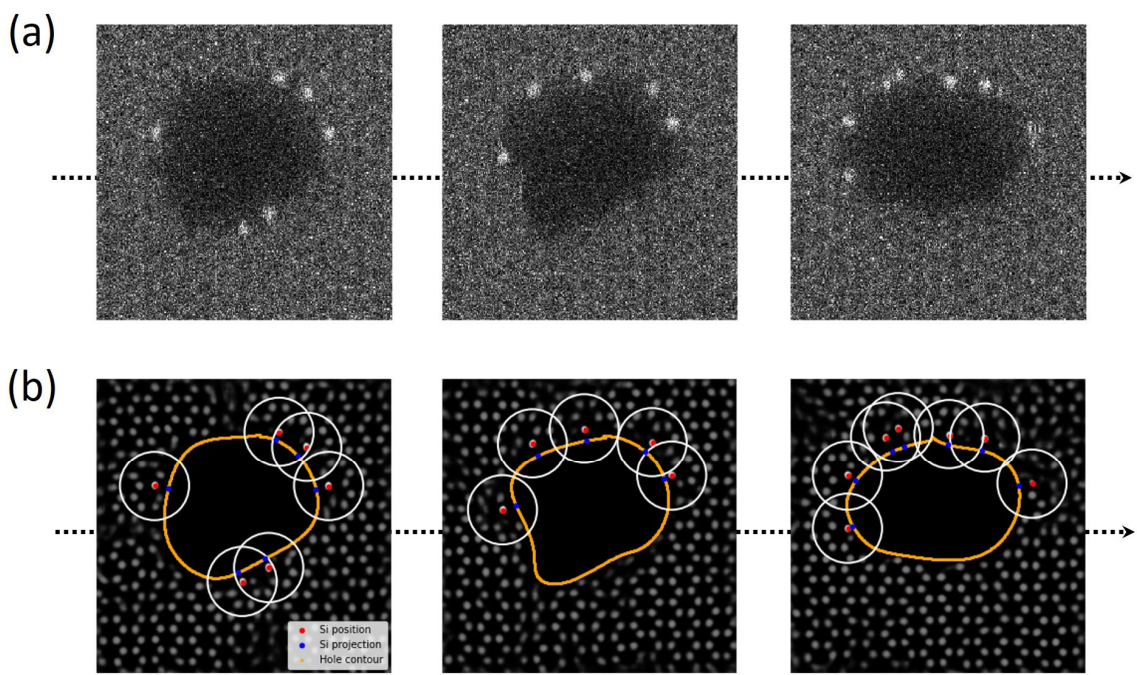

**FIGURE 1.** Experimental STEM data on graphene nanohole with Si impurities. (a) Individual movie frames showing Si impurities (brighter spots) at the edges of nanohole in single layer

graphene. Image size 3 nm × 3 nm. The total number of frames in the movie is 300. (b) Output of deep neural network for data in (a). Red dots denote position of Si atoms. Orange line shows the extracted contour of the nanohole edge for each frame (the nanohole's shape is not constant). Blue dots denote a projection of Si atomic positions on the edge of graphene nanohole. White circles denote sub-images with a local neighborhood of each Si atom that were cropped for further analysis (see text for details).

As a first step of analysis, we employed a fully convolutional deep neural network for image denoising/reconstruction. Our network was inspired by a U-net architecture used for biological and medical images segmentation.[28] The constructed network has a symmetric encoder-decoder structure and utilizes skip connections to provide low-level local features to the high-level global features during upsampling, as well as atrous convolutions in its bottleneck layers to probe features at multiple scales.[29] The models were trained using simulated STEM data[30,31] from three major types of graphene edges, namely, the zigzag, armchair, and bearded edge. Lattice defects such as vacancies and substitutional Si impurities were randomly introduced along the edges. In addition, all the atoms were allowed to randomly move from their initial positions in the ideal lattice by up to 15 % of graphene C-C bond length. The trained model was then applied to experimental data, which allowed us to convert the high-noise video stream into high-contrast data, and determine atomic positions in an automated fashion. In the subsequent analysis, we use both the original data sets, neural network processed data, and atomic coordinates.

Figure 1(a) and Figure 1(b) show the original experimental movie frames and the same frames passed through the trained deep fully convolutional neural network, respectively. The network was able to properly identify location and type of all the atoms in the image, including those at the edge of the graphene hole. In addition, we were able to use the same network for extracting contours of the graphene edge (orange line in Fig. 1(b)).

To analyze the atomistic mechanisms of beam induced transformations, we utilize the Markov model approach. In this approach, the possible (metastable) states of the systems are enumerated. The reaction process is described as memoryless transitions between different states of the system, with the transition probabilities determined only by the initial and final states and not by the prior history or surroundings. The matrix of transition probabilities then fully describes the system dynamics, and enables an analysis of the long-term behavior, presence of dynamic basins, etc. via the structure of the corresponding eigenvalues and eigenvectors.

The mathematical theory and application of Markov processes to physical systems have been well studied for over a century. In the last 20 years, significant effort has been focused on applications of the Markov models for the analysis of molecular dynamic simulation data, as a method to determine the relevant collective slow modes and associated structural elements. Notably, the key element in this analysis was the selection of corresponding descriptors, i.e. the enumeration of possible system states. Here, the projections of the system state on the full parameter space of the system (e.g. dihedral angles) typically gave rise to very high dimensional representations (and hence large Markov matrices), whereas low-dimensional representations typically required considerable domain-specific approximations.

Here, we use the experimental data from the STEM "movie" to create the state descriptors. We used two approaches. In our first approach we tried to apply the multivariate analysis directly to the raw experimental images. For this strategy, we defined the neighborhood of each bright Si atom, and treated the corresponding sub-image (section of the image centered at the atom) as a state descriptor. The additional degree of freedom is the local angle defining the structure orientation in the image plane. Analysis was performed in the absolute coordinate system of the image, which coincides with the graphene sheet, as well as rotation of the coordinate system along the edge of the graphene. However, due to the high noise level in the experimental data, we were unable to automatically detect the details of atomic configurations, except for a few trivial configurations (e.g. one versus three Si atoms), during such analysis.

As a second approach, we used the absolute coordinates and populations of the lattice sites extracted from the data processed by the neural network (i.e. local atomic configurations) as a state descriptor. To obtain data on local atomic configurations around Si impurities, we cropped a circular region of radius equal to ~1.2$a_0$ ($a_0$ - graphene lattice constant) around each Si impurity in every movie frame. The obtained 4D dataset (number of samples × sub-image width × sub-image height × number of channels/classes) was then flattened ("vectorized") and used as an input into a clustering/unmixing algorithm. To estimate the number of components/clusters that our data can be unmixed into and which are related to various Si-C configurations, we explored clusters (dis-)similarity and variability in the dataset (all in terms of image pixels) with hierarchical cluster analysis using Ward's method (Figure 2(a)) and with principal component analysis (Figure 2(b)), respectively. Both methods suggest that one can select the number of components/clusters in the range between approximately 12 and 20. We then performed unmixing for each number of components in that range with three different clustering/unmixing techniques, namely $k$-means clustering, agglomerative clustering and Gaussian mixture model (GMM) methods, which all produced similar results. Here we will only discuss the results of GMM-based unmixing.

The GMM-based unmixing, which allows for rotations, gives rise to large number of components emerging from the fact that, for original data the possible configurations are limited by the symmetry of the original graphene lattice, whereas compensation for local normal gives rise to multiple orientations of the same defects. We therefore have developed a simple procedure based on the structural similarity algorithm[32] to automatically detect and combine classes related by $C_6$ rotation. We further introduced local constraints on the number of atoms within the specific radius around Si atoms (its first coordination sphere) into structure similarity search such that "weak" atoms can also be properly accounted for (above the predefined threshold). We view our approach works as a "recommender system", *i.e.* it advises on which classes are likely related by rotations and thus can be combined, but the final action requires a domain expert approval and also allows for further manual refinement. Interestingly, even after unmixing into a relatively high number of components (e.g. more than 30), this procedure usually allows for quick convergence on the same Si-C configurations (shown with boxes of different color in Fig. 2(c); more details in the next paragraph). In the future, it should be possible to include the discrete symmetry rotations in the deep neural network-based decoding using, for example, equivariant neural networks.[33] The linear unmixing models can also be modified to include the structural sparsity constraints, i.e. allowing only for atomic species with an intensity above a predefined threshold (albeit this approach will exclude dynamic configurations changing faster than frame acquisition times).

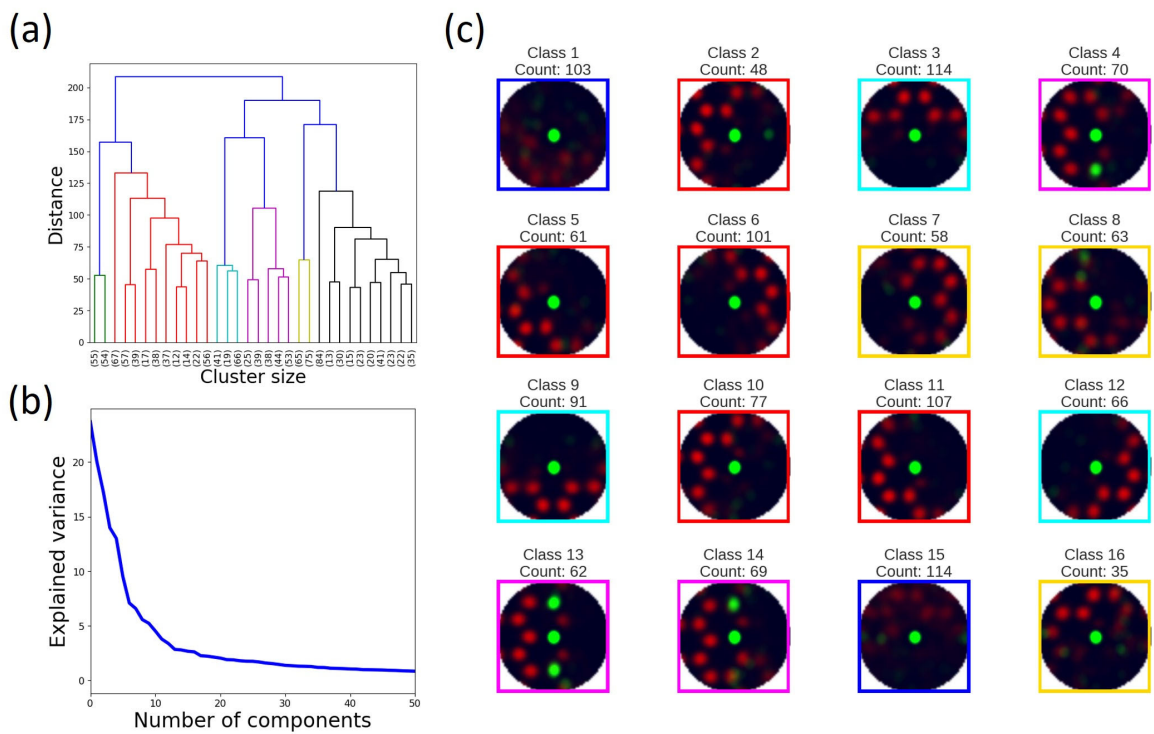

**FIGURE 2.** Statistical analysis of the extracted sub-images centered at Si impurities (see white circles in Fig. 1). (a) Results of the hierarchical cluster analysis (Ward's method). (b) Scree plot of principal component analysis. The plot shows a kink approximately between 12 and 16 components. (c) The result of a Gaussian mixture model-based unmixing of the image stack into 16 components. (c) Results of the hierarchical cluster analysis (Ward's method). The vertical axis shows the distance or dissimilarity between clusters on the horizontal axis. (d) Scree plot of principal component analysis. The kink(s) in the scree plot allows to estimate separation between "signal" (high variance) and "noise" (low variance) in the dataset.

Figure 2c shows the extracted 16 different classes of Si-C configurations, as well as the number of defects in the entire dataset associated with each class. The boxes colored in five different colors show how the classes were grouped using structure similarity search. The green-colored pixels are Si atoms and the dar-red-colored pixels are C atoms. The pixel intensity is associated with the probability of neural network assigning pixels in a specific location to a particular class (in this case, C, Si or background). One can clearly distinguish structures corresponding to i) Si substituting C atom at the zigzag edge apex (blue boxes), ii) Si attached to armchair edge (cyan boxes), iii) Si substituting C atom in the armchair edge (red boxes); iv) Si substituting C atom next to zigzag edge apex atom (yellow boxes), and v) ordering of Si impurities along the zigzag edge (magenta boxes). Notice that in case of Si ordering at zigzag edge the edge apex atoms have weaker intensity suggesting that this class may in fact correspond to a bearded edge (Si atoms attached to zigzag edge apex) with a possibility of reconstruction into zigzag edge configurations (Si atoms located next to edge apex atom) under e-beam irradiation.

The Markov transition matrix derived from frequency of observation of different classes of Si-C edge configurations is shown in Fig. 3. The images of the five associated classes obtained by averaging images in each group shown in Fig. 2(c) are also depicted. The three largest transition probabilities are associated with switching of 1D ordered Si structure into self indicating its relative stability under e-beam, switching between Si attached to armchair edge and Si substituting C at armchair edge, and switching of Si substitution in armchair edge into self (notice that this includes a transition between Si substituting two different C atoms at the edge, which were combined in our analysis and averaged in Fig. 3)

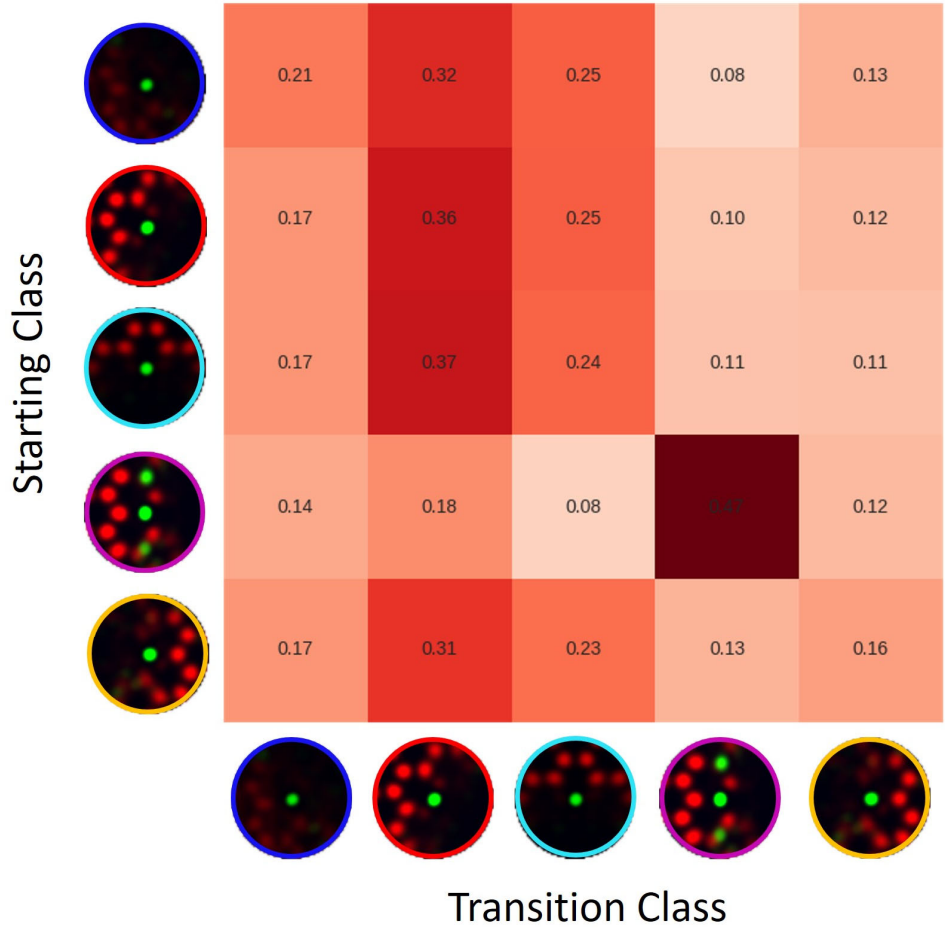

**FIGURE 3. Markov transition matrix for the combined classes of Si-C edge configurations**. The images of the corresponding classes are also shown. These images were obtained by averaging images in each of five different groups shown in Fig. 2(c). Notice that because of the averaging, C atomic features appear on both sides of Si in the second ("red") class in Fig. 3.

An important limitation of the Markov approach is that the processes should be close to reversible, i.e. transition probabilities between states i and j are equal, $\rho_{ij} = \rho_{ji}$. In this case, the Markov matrix is symmetric and corresponding eigenvectors and eigenvalues are real. For experimental data, the matrix can be asymmetric both due to the finite statistics and due to the irreversibility of the process. We note that in this case the observation of long-term dynamics suggest that the process is stationary (the hole does not significantly change size or shape). However, this is the exception rather than the rule for e-beam-induced dynamic processes. Correspondingly, we note that analyses for irreversible processes based on the Koopman operator have been recently proposed.[34] However, we defer these more complex analyses for future studies.

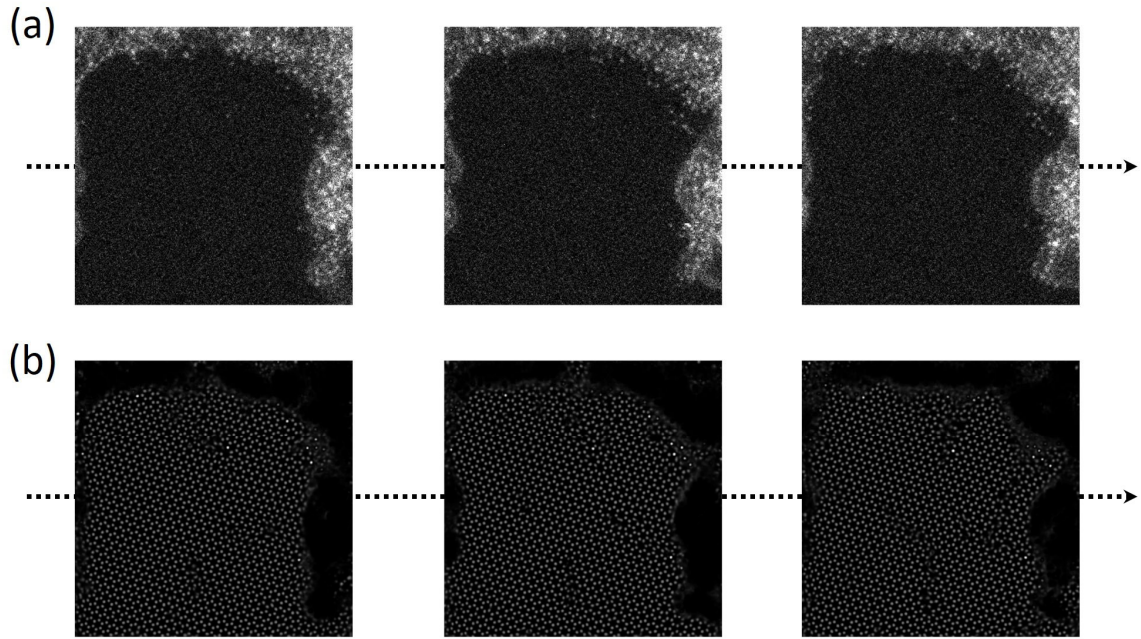

**FIGURE 4.** Experimental STEM data from a region in bulk graphene. (a) Individual movie frames showing noisy graphene lattice, individual Si impurities (brighter spots) and amorphous Si carbon regions (brightest regions). Image size 8 nm × 8 nm. The total number of frames in the movie is 100. (b) Output of deep neural network for data in (a). The network was trained to avoid amorphous Si-C regions as well as individual impurities, which are too close to the amorphous regions.

We now proceed to the analysis of Si transformations in the bulk of graphene. The experimental dataset used for this analysis was obtained from a larger graphene lattice area surrounded by regions of amorphous Si and C as shown in Fig. 4(a). Notice that the lattice structure and Si impurity configurations in the lattice become ill-defined when they are too close to the amorphous regions. We therefore performed an additional (re-)training of our deep neural network to "teach" it how to avoid amorphous Si-C regions as well as Si impurities, which are too close to those regions. The output of the neural network for data in Fig. 4a is shown in Fig. 4b. It is clear that in addition to finding lattice atoms and impurities in noisy data our model can now also easily avoid contaminated regions and impurity atoms that are too close to those regions.

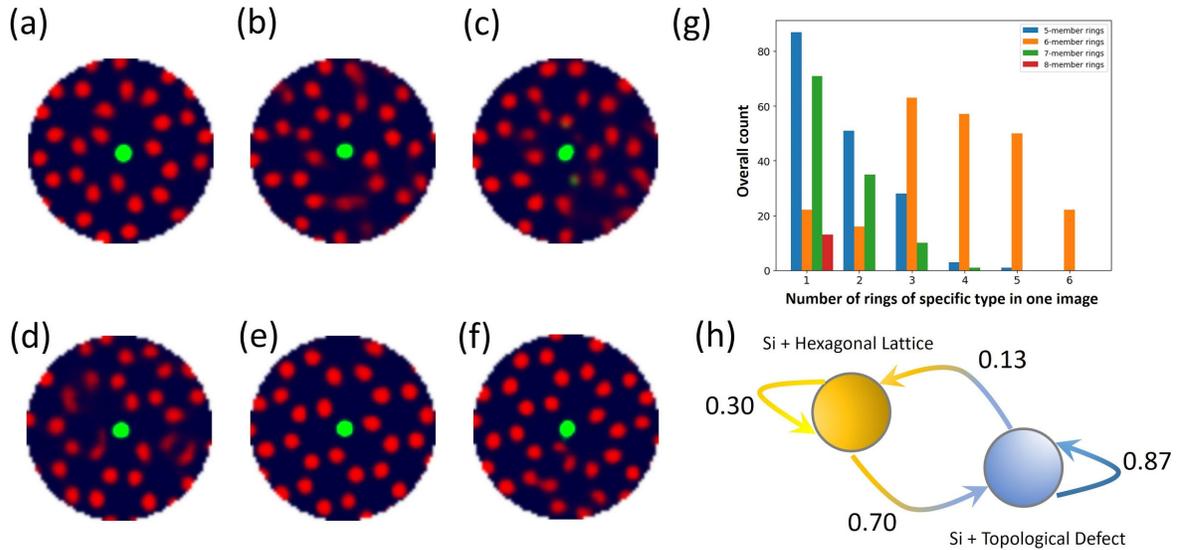

**FIGURE 5.** Analysis of Si impurity configurations in bulk graphene. (a-f) Examples of some configurations of a Si impurity in bulk graphene. In each image, except for (e), the Si impurity couples to topological defects. (g) Statistics of the occurrence of rings of a specific type (e.g. 5-member rings, 6-member rings, etc.). (h) Schematics of Markov transitions between state of Si coupled to topological defect and state without such coupling (surrounded by non-reconstructed hexagons). Transition probabilities are shown next to arrows denoting transitions.

We found that, due to large number of non-hexagonal reconstructions (formation of the so-called topological defects), the previous approach of combining GMM unmixing and structure similarity search based on rotational symmetry does not produce accurate and interpretable results on this dataset. We therefore started with the analysis of graphene "ring network" within ~$2a_0$ around each detected Si impurity utilizing the shortest-path ring search with Franzblau statistics.[35,36] This allowed us to identify the number of 5-, 6-, 7- and 8-member rings present in each cropped region around a Si impurity (Fig. 5(a-f)). Interestingly, we found a large variety of configurations associated with a Si impurity coupling to topological defects. Some examples are shown in Figure 5(a-f). Specifically, Figure 5(a), 5(b) and 5(f) show a defect structure consisting of a Si impurity and 5- and 7-member C rings. Figure 5(d) shows a defect with a Si impurity and 8-member ring, which appears to involve the removal of C atom(s). The previously reported[30,37,38] Si impurity with a 3-fold coordination in non-reconstructed graphene lattice is shown in Figure 5(e). Finally, Figure 5c shows what appears to be a defect corresponding to a partial realization of Si-dimer structure[39] where one of Si atoms "disappeared" during the scan. Two-state Markov analysis and associated transition probabilities for switching between the states of Si impurity with and without coupling to topological defects are shown in Figure 5(h). Notice that here switching from the Si state coupled to a topological defect to itself includes switching events between different topological structures (i.e., different combination and spatial arrangement of 5-, 6-, 7- and 8-member rings). We note that while these defect structures were likely observed in previous studies, the lack of automatic tools for data analysis and defect classification precluded elucidation of these defect classes and analysis of the transformation pathways.

To summarize, we have developed a framework based on the combination of deep neural networks, multivariate statistics, and Markov analysis for the analysis of atomic defect behavior in electron beam induced processes. Specifically, we explored the beam induced reactions of Si atoms on the edge of a graphene nanohole and coupling of Si impurities to topological lattice reconstructions in the bulk of graphene. We believe that this study sets the pathway for the quantitative analysis of elementary mechanisms of solid-state transformations and chemical reactions directly from raw experimental data and can be applied for the analysis of other types of reactions and chemical transformations in different solids on the atomic level.

**Acknowledgements:**

Research was supported by the U.S. Department of Energy, Office of Science, Basic Energy Sciences, Materials Sciences and Engineering Division (SVK). This research was conducted at the Center for Nanophase Materials Sciences, which is a DOE Office of Science User Facility.

**Materials and methods**

Graphene samples were grown via chemical vapor deposition (CVD) on Cu foil and coated with poly(methyl methacrylate) (PMMA) to mechanically stabilize the graphene layer for transfer. The Cu foil was etched away in a bath of ammonium persulfate-deionized (DI) water solution and then rinsed in a bath of DI water to remove the residues of ammonium persulfate. The graphene/PMMA layer was scooped from the DI water bath with a TEM grid (TEMWindows.com product number SN100-A50MP2Q05) and dried for 15 minutes on a hot plate at 150 °C to promote the adherence of the graphene to the TEM grid. After cooling, the PMMA was dissolved by dipping the TEM grid in acetone and gently swishing. While still wet with acetone, the TEM grid was dipped in isopropyl alcohol (IPA) to remove the acetone and allowed to dry. To remove residual contaminants the TEM grid was subsequently baked in an $Ar/O_2$ environment (90%/10%) at 500 °C for 1.5 hours. Prior to loading the samples into the microscope for experimentation, all samples were baked in the sample magazine in vacuum at 160 °C for 8 hours.

**Imaging (experimental conditions)**

The data for the first experiment was acquired using a Nion UltraSTEM U100 operated at an accelerating voltage of 60 kV with a beam current of 60-70 pA. Due to a software bug, metadata for this dataset was not recorded so a precise calculation of dose was not possible. However, based on similar data we estimate a total accumulated electron dose on the order of $10^{10}$ electrons/nm$^2$. Data for the second experiment was acquired using a Nion UltraSTEM U200 operated at an accelerating voltage of 100 kV with a beam current of 10-20 pA. Using 20 pA we calculate a total accumulated dose of $8.18 \times 10^8$ electrons/nm$^2$. All images were acquired with the high angle annular dark field (HAADF) detector.

**Data analysis**

Deep neural networks were trained with either Tensor Processing Unit (TPUv2) or with Tesla K80 Graphical Processing Unit (GPU) in Google Collaboratory with Tensorflow and Keras deep learning libraries. All the data analysis is available in a form of Jupyter notebook at https://github.com/ziatdinovmax/Notebooks-for-papers/blob/d784d48add1f90480ffcfc648624a5adb2e23007/Si-atom-dynamics-in-graphene.ipynb